\documentstyle[twoside,fleqn,espcrc2,epsf,rotate]{article}

  \newcommand{\cZ}{\cal{Z}}

\newcommand{\AmS}{{\protect\the\textfont2
  A\kern-.1667em\lower.5ex\hbox{M}\kern-.125emS}}

\title{Phase Structure of Lattice QCD at Finite Density with Dynamical Fermions }
\author{
I.M. Barbour\address{Dept. of Physics and Astronomy,University of Glasgow G12 8QQ, U.K., (UKQCD Collaboration)},
J.B. Kogut\address{Dept. of Physics,University of Illinois, 1110 West Green Street,Urbana, IL 61801}
and 
S.E. Morrison$^{{\footnotesize{a}}}$
\thanks{talk presented by S.E. Morrison}
}       
\begin{document}
\begin{abstract}
We compare the chemical
potential associated with the onset of non-zero baryon number 
density on $6^4$ and $8^4$ lattices at $\beta=5.1$ and ma=0.01. 
We provide evidence for $Z(3)$ tunnelling. We determine a 
critical chemical potential of $\mu a \simeq 0.1$ which is 
unexpectedly low. 
We also determine
the dependence of the onset of the observed phase transition on 
the quark mass. The physically misleading result of the quenched 
theory is shown to persist despite the inclusion of the complex fermion determinant.
\end{abstract}
\maketitle
\section{Introduction}
   Lattice QCD at finite baryon density holds the key to an understanding of
the phase transition between quark-confined hadronic matter 
and the quark-gluon plasma. 
At zero temperature we expect this 
transition to occur at $\mu = { {m_p}\over 3}$ 
corresponding to the lowest lying state with
non-zero baryon number. \\
	The fact that the fermion determinant becomes complex when $\mu$ is non-zero  makes numerical simulations particularly difficult. Simulations using the quenched theory \cite{BBDKMSW86} give the physically misleading result that as the quark mass tends to zero, the lowest lying state with non-zero baryon number becomes massless. This corresponds to
${\mu}_c \simeq{ {m_{\pi}}\over {2}}$. Recent work by Stephanov \cite{Steph96} using the random matrix method has 
provided a solution to this particular problem by demonstrating that at nonzero $\mu$ the quenched theory is not a simple $n \rightarrow 0$ limit of a theory with $n$ quarks. This explanation implies that the phase of the determinant is important for a simulation of full QCD at finite
$\mu$. However earlier work on the full theory at strong coupling \cite{Lombardo}  indicates that the transition
still occurs at ${ {m_{\pi}}\over {2}}$.
\subsection{Formulation}
The GCPF can be expressed as an ensemble average of the fermionic determinant
normalised with respect to the fermionic determinant at $\mu=0$. This 
means that our probability weighting factor is well-defined.

\begin{equation}
{\cZ}={{\int [dU][dU^\dagger]det M(\mu)e^{-S_g[U,U^\dagger]}}\over {
\int [dU][dU^\dagger]det M(\mu=0)e^{-S_g[U,U^\dagger]}}}
\end{equation}
which leads to:
\begin{equation}
{\cZ}=\left<{{detM(\mu)}\over {detM(\mu=0)}}\right>_{\mu=0}
\end{equation}
where 
$S_g$ is the pure
gauge action and $M$ is  the Kogut-Susskind fermion matrix:
\begin{equation}
iM=G+ im +Ve^{\mu}+V^{\dagger} e^{-\mu}.
\end{equation}
 
$G$ contains all the spacelike gauge links and the bare quark mass and $V$ all the forward timelike links.
On each given configuration of the gauge fields
$det M$ is expanded as a polynomial
in the fugacity variable with powers ranging between 
$[-3n_s^3,3n_s^3]$ ($n_s$ corresponding to
the spatial extent of the lattice).
The ensemble averages of the coefficients of this normalised expansion 
are the corresponding expansion coefficients for the GCPF.

The gauge field configurations are generated using Hybrid Monte
Carlo, 
with the expansion coefficients of $det M$ obtained from
the eigenvalues of the propagator matrix $P$ (Gibbs \cite{Gibbs86})
\begin{equation}
P=\left(\begin{array}{cc}
-GV & V \\
-V  & 0
\end{array} \right)
\end{equation}
 Thus (for $a=1$):
\begin{equation}
\det(iM) = e^{3\mu n_s^3 n_t} \det(P-e^{-\mu}).
\end{equation}
\begin{equation}
det(P-e^{-\mu})=\sum_{n=0}^{6 n_s^3 n_t}w_ne^{-n\mu} 
\end{equation}

The eigenvalues, $\lambda$, of $P$ have a $Z(n_t)$ symmetry and a $\lambda$ and $1/ \lambda^{*}$ symmetry, which  
leads to an expansion of the form:
\begin{equation}
{\cZ}=\sum_{n=-3{n_s}^3 }^{3{n_s}^3}<b_{|n|}>e^{n\mu {n_t}}
=\sum_{n=-3{n_s}^3}^{3{n_s}^3}e^{-(\epsilon_n - n\mu)/T}
\end{equation}

\subsection{$Z(3)$ Tunnelling}
In the pure gauge theory we have $N_c$ equivalent vacua related by $Z(N_c)$ rotations. Tunnelling between the different $Z(3)$ vacua is much more probable
in the confined sector than in the deconfined sector. Since the pure-gauge action as well as the integration measure are invariant under the $Z(3)$
transformation, the GCPF can also be written as:
\begin{equation}
{\cZ}(\mu)={{\int [dU][dU^\dagger]det M(\mu+z_3/n_t)e^{-S_g[U,U^\dagger]}}\over {
\int [dU][dU^\dagger]det M(\mu=0)e^{-S_g[U,U^\dagger]}}}
\end{equation}
Averaging over the three $Z(3)$ vacua
would eliminate the triality non-zero coefficients and, since $Z(0)=1$, the sum of the triality zero
coefficients should tend to 1.
As Fig. 1 shows,at $\beta=5.1$ and $m=0.01$, we obtained a clear signal for Z(3) tunnelling. 
The sum of the triality zero coefficients shows a strong tendency to 
average to 1 whilst those of the triality
one and two coefficients each tend to zero. This  is consistent with
simulating in the confined phase at $\mu=0$ as expected at this coupling and quark mass.

\begin{figure}[htb]
\vspace{-0.6cm}
\epsfxsize=5.5cm
\rotate[r]
{\epsfbox{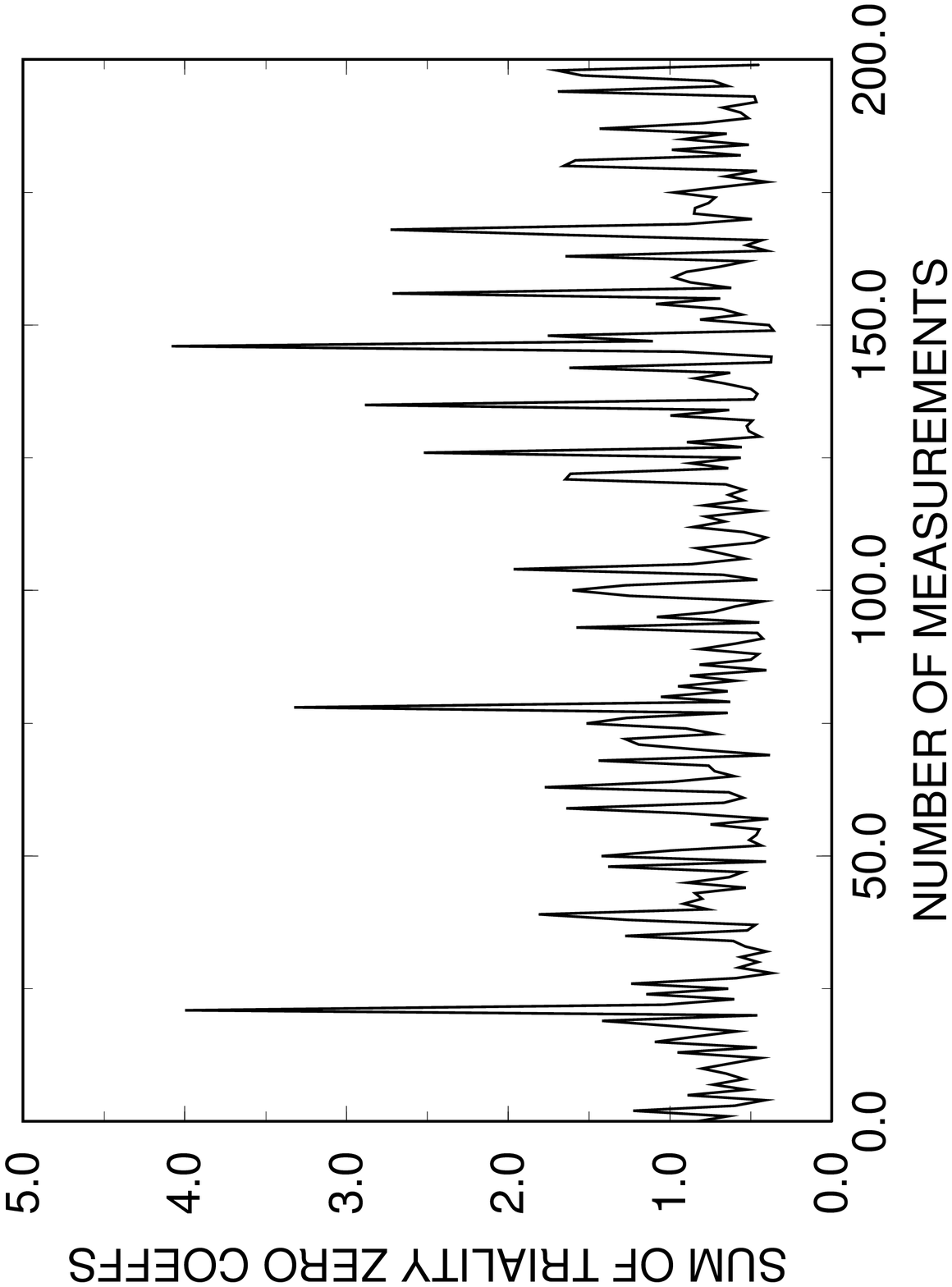}}
\vspace{0.2cm}
\rotate[r]
{\epsfbox{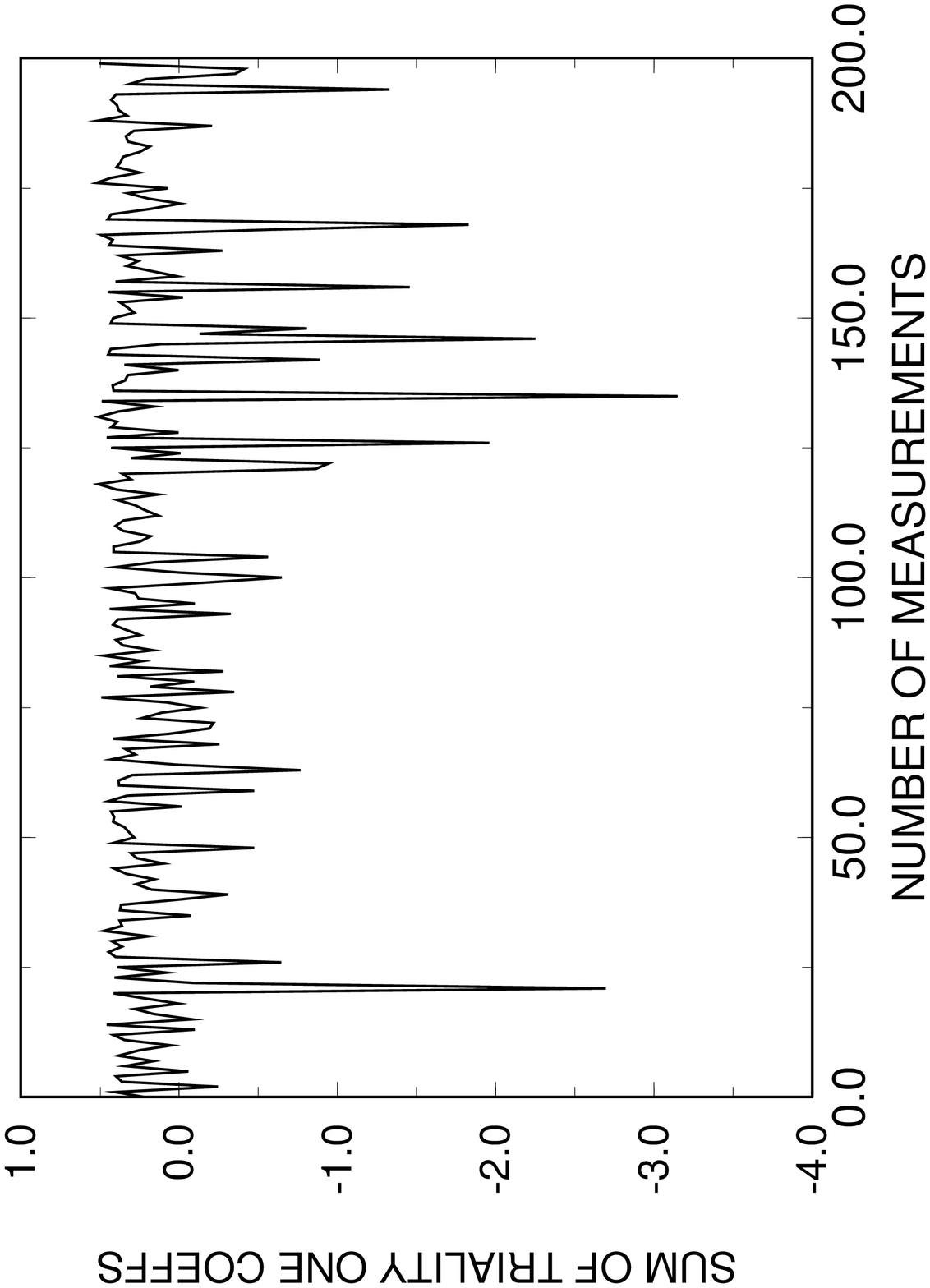}}
\vspace{0.2cm}
\vspace{-0.6cm}
\caption{The HMC time evolution of the expansion coefficients on an $8^4$ lattice 
at $\beta=5.1$ and $ma=0.01$. The behaviour of the triality 2 coefficients is similar to that
of the triality 1 coefficients.}
\vspace{-0.6cm}
 \end{figure}

\subsection{Baryon Number Density}

	 We consider a description of the system in terms of  the canonical partition 
functions for fixed particle number. This technique allows us to
include only those triality zero coefficients which, at the end of the simulation, are positive and have a reasonable error.The chemical potential as a function of the baryon number density 
is obtained from the local derivative 
of the energy, $\epsilon_n$, of the state with $n$ fermions with respect to $n$. 
\begin {equation}
\mu(\rho) = {{1}\over {3 n_s^3 }}{{\partial \epsilon_n}\over{\partial\rho}}
\end {equation}
where $\rho$ is the fermion density, $n\over{3n_s^3}$.

\begin{figure}[htb]
\epsfxsize=5.5cm
\rotate[r]
{\epsfbox{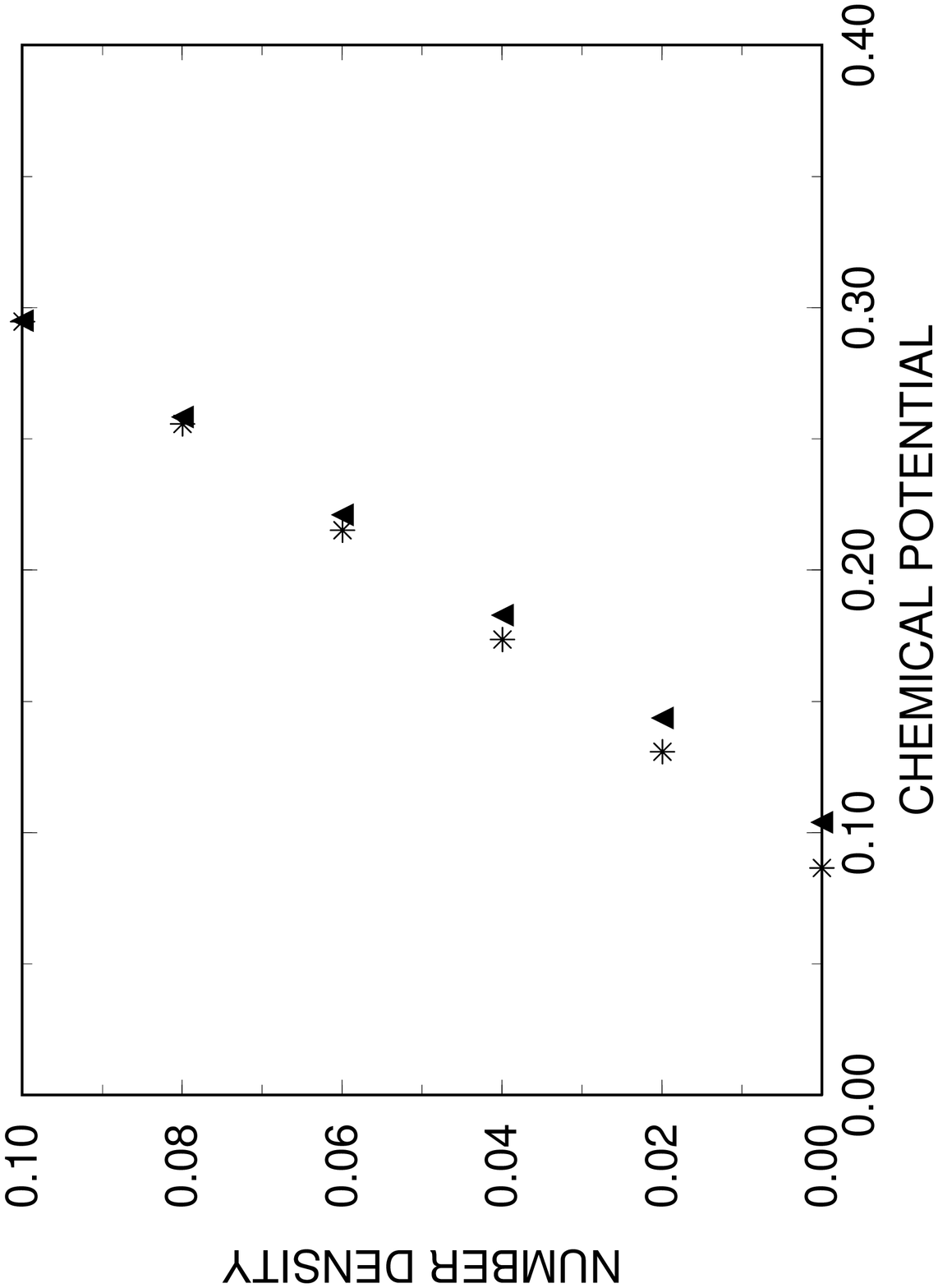}}
\vspace{-0.6cm}
\caption{The fermion number density \newline on $6^4$ (triangles) and $8^4$ (stars) lattices at \newline $\beta=5.1,ma=0.01$. }
\vspace{-0.6cm}
 \end{figure}
		
	In Fig. 2 we show the number density, at $\beta=5.1$, $m=0.01$ on $6^4$ and 
$8^4$ lattices, as a function of chemical potential.
Only the dependence at low densities is shown as it 
is only the coefficients $<b_n>$ for $n$ small
which are determined with reasonable statistical error.
There is little difference between the $6^4$ and the $8^4$ number densities.
The onset of non-zero baryon number density at ${\mu}_c
\simeq 0.1$ is unexpectedly low.

\subsection{Mass Dependence of the Onset $\mu$}

  Therefore we have also considered the chemical potential (the onset $\mu$) required to 
make the level with 3 quarks  equally probable with the zero quark level. 
This ad-hoc definition takes account of the fact that the lowest coefficients are the most accurately determined and allows errors to be estimated directly. 
	Fig. 3 shows the quark mass dependence of the onset $\mu$. Note that, from a 
study of the Lee-Yang zeros in the complex m-plane, we expect the system
to undergo a `deconfinement transition' in the region $m<0.01$ on a $6^4$ lattice at $\beta=5.1$. 
The scaling with $\sqrt m$ is clearly preferred, consistent with the onset of the transition
being controlled, in part, by a Goldstone boson.

\begin{figure}[htb]
\epsfxsize=5.5cm
\rotate[r]
{\epsfbox{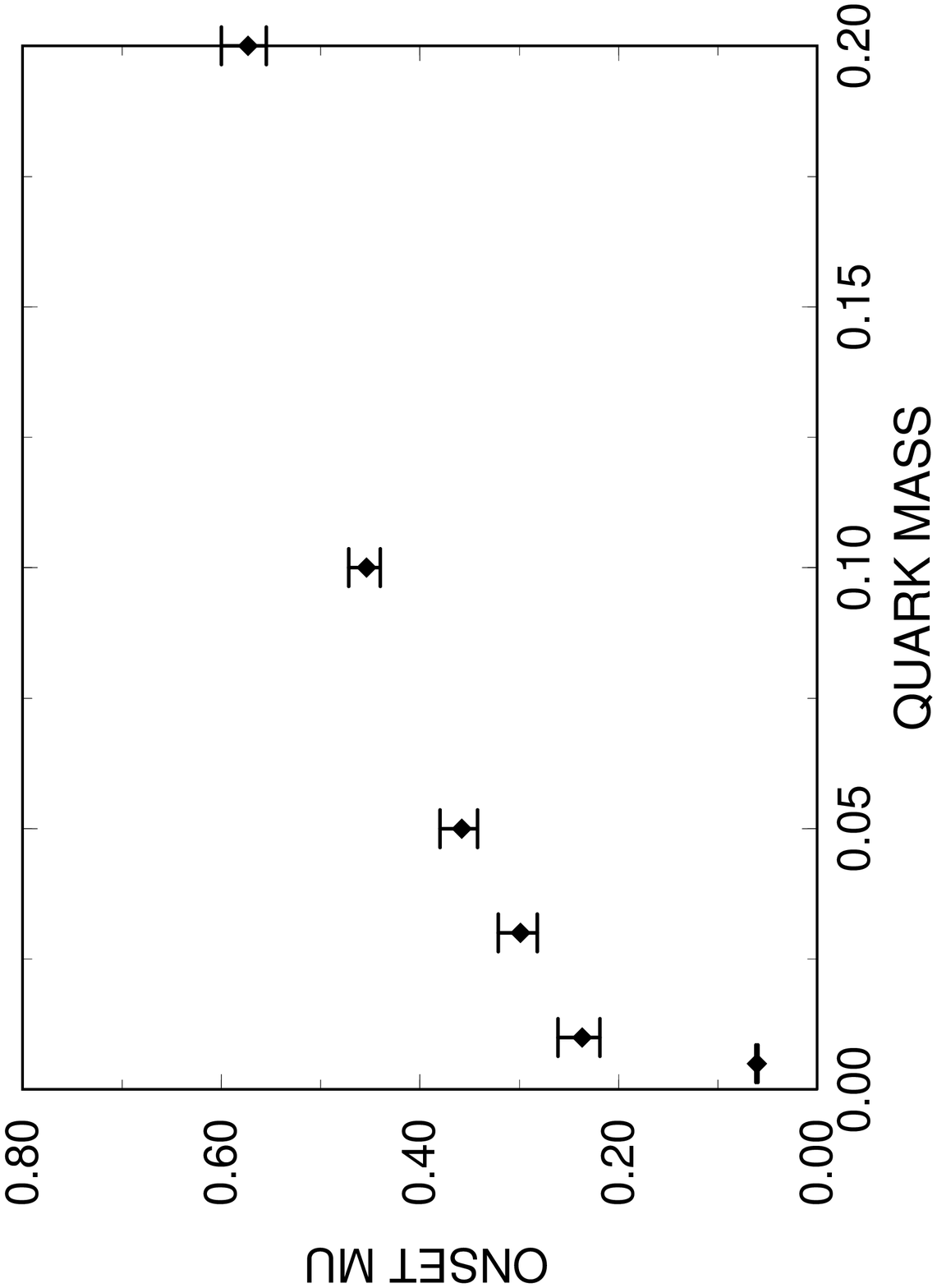}}
\vspace{0.2cm}
\rotate[r]
{\epsfbox{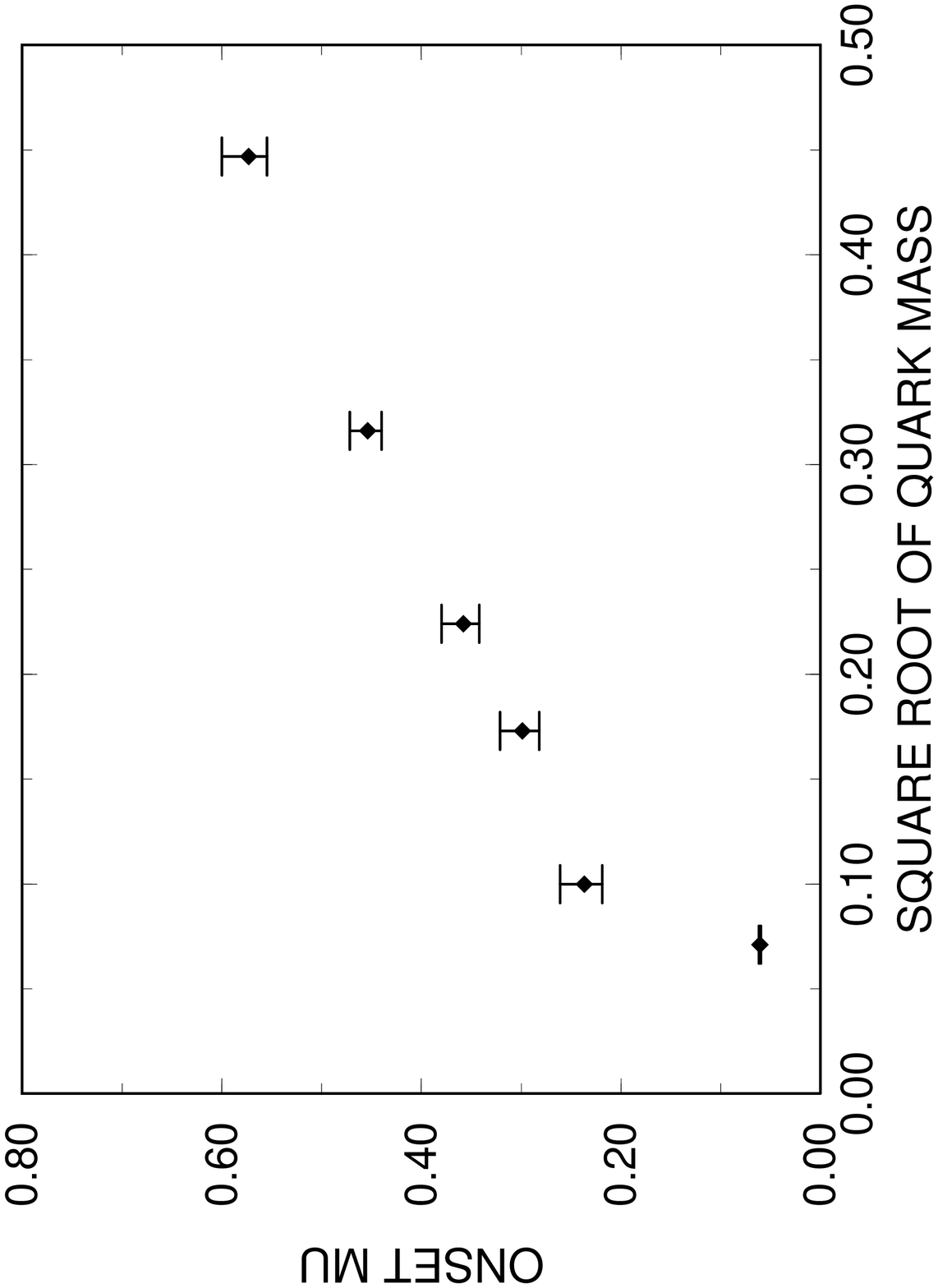}}
\vspace{-0.6cm}
\caption{The quark mass dependence of the onset $\mu$ ($\beta=5.1$ on a $6^4$ lattice). }
\vspace{-0.6cm}
 \end{figure}

\section{Conclusions}
	Lattice QCD at finite baryon density continues to exhibit a phase 
transition inconsistent with ${\mu}_c \simeq {{m_p}\over{3}}$ despite the inclusion
of the complex fermion determinant. It may well be that a successful formulation of QCD
at finite density  must
override the effects of a Goldstone boson.   

\section{Acknowledgements}
        We would like to thank Misha Stephanov, Maria-Paola Lombardo
and Ely Klepfish for very useful suggestions. This work was supported 
in part by the NSF through grant NSF-PHY92-00148,by DOE, by 
PPARC through grant GR/K55554 and by NATO via grant CRG960002.
Computations were performed on the C90s at NERSC and PSC.

\

\end{document}